\documentclass[twocolumn,english,aps,prb,floatfix,preprintnumbers,showpacs,amsfonts,amssymb,superscriptaddress]{revtex4}
\usepackage{ae,aecompl}
\usepackage[T1]{fontenc}
\usepackage[latin1]{inputenc}
\usepackage{color}
\usepackage{array}
\usepackage{amsmath}
\usepackage{amssymb}
\usepackage{graphicx}

\makeatletter

\providecommand{\tabularnewline}{\\}

\@ifundefined{textcolor}{}
{%
 \definecolor{BLACK}{gray}{0}
 \definecolor{WHITE}{gray}{1}
 \definecolor{RED}{rgb}{1,0,0}
 \definecolor{GREEN}{rgb}{0,1,0}
 \definecolor{BLUE}{rgb}{0,0,1}
 \definecolor{CYAN}{cmyk}{1,0,0,0}
 \definecolor{MAGENTA}{cmyk}{0,1,0,0}
 \definecolor{YELLOW}{cmyk}{0,0,1,0}
}

\usepackage{amscd}
\usepackage{bm}
\usepackage{graphics,psfrag}
\usepackage{graphicx,psfrag}
\usepackage{lipsum}

\makeatother

\usepackage{babel}
\begin{document}

\title{The $N$-Leg spin-$S$ Heisenberg ladders: A DMRG study}

\author{F.~B.~Ramos}

\affiliation{Universidade Federal de Uberlândia, Instituto de F\'{\i}sica, Caixa
Postal 593, 38400-902 Uberlândia, MG, Brazil }

\author{J.~C.~Xavier}

\affiliation{Universidade Federal de Uberlândia, Instituto de F\'{\i}sica, Caixa
Postal 593, 38400-902 Uberlândia, MG, Brazil }

\date{\today{}}
\begin{abstract}
We investigate the $N$-leg spin-$S$ Heisenberg ladders by using
the density matrix renormalization group method. We present estimates
of the spin gap $\Delta_{s}$ and of the ground state energy per site
$e_{\infty}^{N}$ in the thermodynamic limit for ladders with widths
up to six legs and spin $S\leq\frac{5}{2}$. We also estimate the
ground state energy per site $e_{\infty}^{2D}$ for the infinite two-dimensional
spin-$S$ Heisenberg model. Our results support that for ladders with
semi-integer spins the spin excitation is gapless for $N$ odd and
gapped for $N$ even. Whereas for integer spin ladders the spin gap
is nonzero, independent of the number of legs. Those results agree
with the well known conjectures of Haldane and Sénéchal-Sierra for
chains and ladders, respectively. We also observe edge states for 
ladders with $N$ odd, similar to what happens in
spin chains.
\end{abstract}


\maketitle

\section{Introduction}

The theoretical study of strongly correlated systems is undoubtedly
an extremely complicated task due to the lack of appropriate techniques
to address such systems, especially in dimensions greater than one.
The main difficulty in the investigation of those systems is associated
to the fact that the Hilbert space grows exponentially with the system
size. In the last decades a great deal of effort has been devoted
to develop new techniques for dealing with this issue. A major breakthrough,
in this area, was the development of the density matrix renormalization
group (DMRG) by White.\citep{white} The success of the DMRG resides
in the choose of the optimal states used to represent a Hamiltonian,
which are selected through the reduced density matrix eigenvalues.\citep{white}
Usually, only a small percentage of the whole Hilbert space is necessary
to describe the low energy physics of a system by using the DMRG.
Due to this fact the DMRG became one of the most powerful techniques
to deal with strongly correlated systems in one-dimension. Although
the DMRG is based on a one-dimensional algorithm, it has been applied
to low dimensional systems such as the ladder systems.\citep{dagottosc1}
The procedure consists in mapping the low-dimensional model on a 1D
model with long range interactions.\citep{dmrg2dliang,whitenleg,dmrg2dwhite,Henelius2d,dmrg2dxiang,xavierqcp,dmrg2dfarnell,xiang-nleg,nishimoto3leg,capponi3leg,whitereview2d,caponi-nleg,nishimoto-kagome}
In this vein some other algorithms, based in the tensor networks,
have been proposed to study strongly correlated systems in
dimensions higher than $d=1$, such as PEPS\citep{verstraetePEPS} and MERA.\citep{Vidalmera}
Note that in dimensions $d>1$ there are few accurate results which
can be used as benchmark. 

\begin{figure}[!]
\begin{centering}
\label{fig:sch-rep}
\par\end{centering}

\begin{centering}
\includegraphics[width=6cm]{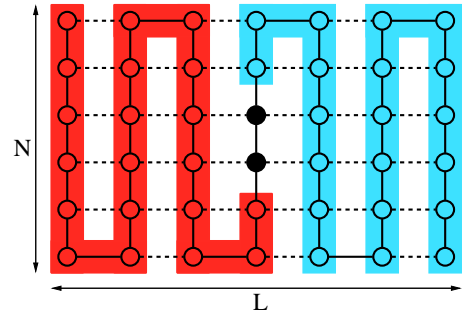}
\par\end{centering}

\caption{(color online) Schematic representation of a $N$-leg ladder composed
of $N$ chains of size $L$ from the point of view of the DMRG. The
circles represent the sites. The solid (dashed) lines represent nearest-neighbors
(long-range) interactions. The sites covered by red (blue) correspond
to the block $A$ ($B$) in the context of the DMRG. The filled (open)
circles are the center sites (renormalized sites).}
\end{figure}

One of the goals of this work is to provide accurate results of the
$N$-leg spin-$S$ Heisenberg ladders, with spin up to $S=\frac{5}{2}$,
which may be used as benchmark for the new algorithms that have been
proposed. Besides, and not less important, we intent to verify the
Haldane-Sénéchal-Sierra conjecture,\citep{haldaneconj1,haldaneconj2,senechal-nleg,Sierralegs}
which dictates the behavior of the spin gap of the spin-$S$ Heisenberg
chains and ladders.

The $N$-leg ladders are characterized by $N$ parallel chains coupled
one to each others, such that the coupling $J$ along the chains is
comparable to the coupling $J_{\perp}$ between the chains (see Fig.
1). The $N$-leg ladders are easier to deal with than the two-dimensional
systems and are used as an simple route to study the last
ones. \citep{dagottosc1} In this work, we focus in legs formed
by spin-$S$ Heisenberg chains. It is well known that the spin-$S$
Heisenberg chains, i. e. $N=1$, present a very distinct behavior
with the value of $S$, as first pointed out by Haldane. \citep{haldaneconj1,haldaneconj2}
Haldane by using a semi-classical limit of the Heisenberg chain noted
that semi-integer (integer) spin Heisenberg chains are gapless (gapped),
this latter statement is known as Haldane conjecture. Although the
semi-classical approach is valid for $S\gg1$ the conjecture was verified
numerically for $S\ge1/2$.\citep{moreo-chains,liang-chains,spinschico}
In the context of the ladder systems, the same natural question arises.
Dagotto \textit{et al.}, in Ref \onlinecite{dagotto2leg}, were the
first to show that the two-leg spin-$\frac{1}{2}$ Heisenberg ladder
is gapped. In Ref. \onlinecite{rice-nleg} Rice \textit{et al.} argued
that spin-1/2 Heisenberg ladders with an odd (even) number of legs
are gapless (gapped). Indeed this was verified numerically with DMRG\citep{whitenleg}
(see also Ref. \onlinecite{troyer-nleg}). A similar analysis as the one done
by Haldane for the spin-$S$ Heisenberg chains was extended for the
Heisenberg ladders by Sénéchal in Ref. \onlinecite{senechal-nleg}
and by Sierra in Ref. \onlinecite{Sierralegs}. These authors concluded
that the $N$-leg spin-$S$ Heisenberg ladders is gapless (gapped)
if $SN$ is semi-integer (integer). While there are strong numerical
evidences that the Haldane-Sénéchal-Sierra conjecture hold for ladders
with spin $S=1/2$ (see Refs. \onlinecite{whitenleg} and \onlinecite{troyer-nleg})
very few works consider ladders with $S>1/2$. In particular, the
$N$-leg spin-1 Heisenberg ladder was studied by bozonization,\citep{senechal-nleg-bozo,antuso-nleg-per-bozo}
Monte Carlo\citep{todo-n1-s1},  perturbation theory methods\citep{sato-nleg-pert}
and by using the non-linear sigma model
approach\citep{senechal-nleg,Sierralegs,satonlsm} 
(see also Ref. \onlinecite{capponi3leg}). 

The study of quantum ladders is not just a theoretical artifact to reach the
two-dimensional systems. Compounds such as vanadyl pyrophosphate $\mathrm{(VO_{2})P_{2}O_{7}}$
and some cuprate systems, such as $\mathrm{SrCu_{2}O_{3}}$ and $\mathrm{Sr_{2}Cu_{3}O_{5}}$,
are examples of experimental realizations of the spin-1/2 Heisenberg
ladders.\citep{dagottosc1} Certainly the study of ladders with larger
spins are highly desired, not only from the theoretical point of view.
Because there are compounds, such as $Na_{2}Ni_{2}(C_{2}O_{4})_{3}(H_{2}O)_{2}$
%Since exist compounds, such as $Na_{2}Ni_{2}(C_{2}O_{4})_{3}(H_{2}O)_{2}$
and $\beta-CaCr_{2}O_{4}$ that are experimental realizations of ladders
with spin-1\citep{mennerich-2leg-s1} and spin-3/2,\citep{damay-s32-2010,damay-s32}
respectively. 

\begin{figure}[!]
\begin{centering}
\label{fig:sch-rep-1}
\par\end{centering}

\begin{centering}
\includegraphics[width=9cm]{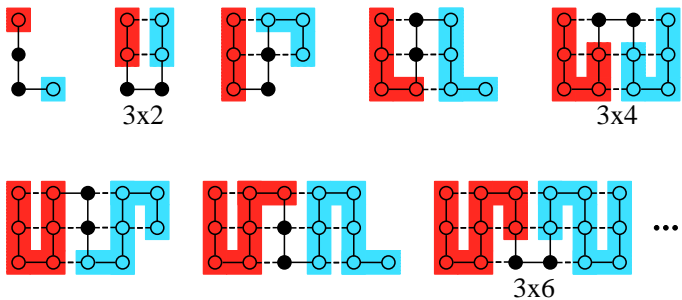}
\par\end{centering}

\caption{(color online) Illustration of the growth of a 3-leg ladder in the
infinite system DMRG algorithm. The notation of the symbols is the
same of Fig. 1.}
\end{figure}

In this work, we consider the Hamiltonian of the $N$-leg spin-$S$
Heisenberg ladders, under open boundary conditions, defined by

\begin{equation}
H=J\sum_{i=1}^{N}\sum_{j=1}^{L-1}\mathbf{S}_{i,j}\cdot\mathbf{S}_{i,j+1}+J\sum_{i=1}^{N-1}\sum_{j=1}^{L}\mathbf{S}_{i,j}\cdot\mathbf{S}_{i+1,j},\label{eq:hamiltonian}
\end{equation}

\noindent where $\mathbf{S}_{i,j}$ is the spin-$S$ operator at the
$i$-th leg and $j$-th rung, $L$ is the number of sites of the chains
and $N$ is the number of legs. We have set $J=1$ to fix the energy
scale.

\noindent We use the DMRG algorithm to investigate the Hamiltonian
above, keeping up to $m=3600$ states per block in the final DMRG
sweeps and the discarded weight was typically $10^{-8}-10^{-12}$
at the final sweeps. In order to avoid metastable configurations we
start the truncation process with large values of $m$ (typically
we start with $m_{0}\sim1200$).\citep{xavieravoid}

In the next section we present our results for the spin-$S$ $N$-leg
Heisenberg ladders: estimates of the ground state energy per site
$(e_{\infty}^{N})$ as well the estimates of the spin gap $(\Delta_{s})$
in the thermodynamic limit. We also show estimates of the ground state
energy per site of the two-dimensional spin-$S$ Heisenberg model.
A discussion of an edge effect, similar to what happens in the spin-1
chain, is also reported for ladders.
% with integer spins.

Before we start presenting our results, let us briefly describe the
procedure we used to calculate the energies of ladder systems by using
the DMRG. As we already mentioned before, the DMRG algorithm is essentially
a 1D algorithm. However, it is possible to map the cluster of size $N\times L$
in a one-dimensional system with long-range interactions, as illustrated
in Fig. 1. Before being reached a specific cluster size (as the one
presented in Fig. 1) we need to grow the size of the system starting with
four sites. In the infinite system DMRG algorithm at each step two sites are added,
as illustrated in the Fig. 2 for the case
of a 3-leg ladder. It is very interesting to note that, as in the one-dimensional
case,\citep{white} we can also use the infinite system DMRG algorithm
to obtain  the ground state energy per site, in the thermodynamic,
$e_{\infty}^{N}$ of the $N$-leg ladders. The energy per site can
be estimated from the difference in energy of different iterations.
Note that some energies (associated with some iteractions) can
not be used to estimate $e_{\infty}^{N}$, since not all steps have
the correct number of sites for a fixed geometry (see Fig. 2). We estimate $e_{\infty}^{N}$
by using the following equation

\begin{equation}
e_{\infty}^{N}=\lim_{L\rightarrow\infty}\frac{E\left[N(L+2)\right]-E\left(NL\right)}{2N},\label{eq:e0-geral}
\end{equation}
where $E(M)$ is the ground state energy of a system with $M$ sites.
To our knowledge this simple procedure to estimate $e_{\infty}^{N}$
for ladder systems was not used in the literature before.

\section{Results}

\subsection*{The ground state energy per site: $e_{\infty}^{N}$ }

In Table I, we present our accurate estimates of $e_{\infty}^{N}$
for several values of $N$ and $S$. In order to get these estimates
we first used Eq. (\ref{eq:e0-geral}) for a fixed number of states kept
in the truncation process $(m$) and increase $L$ until the results
converge (for large values of $m$ and $N$ the number of iterations
is of the order of one thousand). We then get another estimate with
a larger value of $m$ (typically two times larger) and we compare
this estimate with the previous one. The number of digits shown
in this table corresponds to the precision we get considering $m$
up to 1600. The results presented in parentheses are some of the best
estimates known in the literature 
(see also Refs. \onlinecite{barnerladders,qinspin-1-2,golinelli-spin-1,Sunspin2,qin-spin2b,peterentropy}
for similar estimates). As we can see, our estimates are in perfect
agreement with those results. 

Since we were able to get accurate estimates of $e_{\infty}^{N}$
for legs up to $N=6$ we decide to estimate the ground state energy
per site $e_{\infty}^{2D}$ of the infinite two-dimensional system
by assuming that $e^{2D}(N,L=\infty)=e_{\infty}^{N}$ behaves as

\begin{equation}
e_{\infty}^{N}=e_{\infty}^{2D}+\frac{A}{N}.\label{eq:fit-inf}
\end{equation}
The above behavior is expected even for non-interacting systems,
as shown below. Consider the following non-interacting Hamiltonian

\[
H=-\sum_{<i,j>,\sigma}(c_{i,\sigma}^{\dagger}c_{j,\sigma}^{\phantom{\dagger}}+\mathrm{H.c.})\,,
\]
where $c_{j\sigma}$ annihilates a  electron at site $j$
with spin projection $\sigma$. Here $\langle ij\rangle$ denotes
nearest-neighbor sites and we are considering periodic (open) boundary
condition in the direction x (y). Let $L$ $(N$) be the number of
sites in the x (y) direction. It is not difficult to shown 
that the ground state energy at the half-filling 
for $L\rightarrow\infty$ is given
%For $L\rightarrow\infty$ is not difficult
%to show that the ground state energy at the half-filling is given
by 
\[
\lim_{L\rightarrow\infty}\frac{E(N,L)}{L}=-4\sum_{j=1}^{N}\Bigg[ \frac{1}{\pi}\sin\left(\frac{N-j+1}{N+1}\right)+
\]
\[
+\frac{N-j+1}{N+1}\cos\left(j\pi/(N+1)\right)\Bigg].
\]
 The leading finite-size corrections of the above equation can be
obtained by using the Euler-Maclaruin formula (see for example
Ref. \onlinecite{eloyxavier} for a similar case of a two-leg spin ladder). The result
obtained is

\[
\lim_{L\rightarrow\infty}\frac{E(N,L)}{NL}=-\frac{16}{\pi^{2}}+(2-16/\pi^{2})\frac{1}{N}\ ,
\]
which presents a scaling form similar to one of  Eq.  (\ref{eq:fit-inf}).

In Fig. 3, we show $\frac{e_{\infty}^{N}}{S^{2}}$ as function of
$\frac{1}{N}$ for several values of spins. The symbols in this figure
are the numerical data and the dashed lines connect the fitted points
using Eq. (\ref{eq:fit-inf}). As we can observe from the figure,
the fits are very good. The values of $e_{\infty}^{2D}$ acquired
from the fits are also listed in Table I. As we can see in this table,
our estimate for the case $S=1/2$ ($e_{\infty}^{2D}=-0.6768$) agrees
with the one obtained by Monte Carlo method ($e_{\infty}^{2D}=-0.66931$).\citep{Wiesesspin1-2-2d} 
The origin of the very small difference (0.007) between these two values 
is very probably associated  with  the small lattice sizes considered to extrapolated
our data.

\begin{figure}
\begin{centering}
\includegraphics[width=7cm]{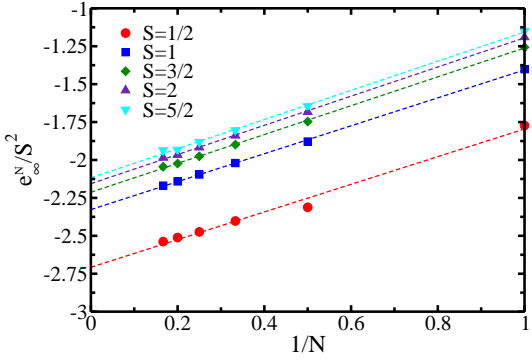}
\par\end{centering}

\caption{(color online) Ground state energy per site of infinite N-leg Heisenberg
ladders $e_{\infty}^{N}/S^2$ vs $\frac{1}{N}$ for spins up to S=$\frac{5}{2}$
(see legend). The dashed lines are fits to our data using Eq. (\ref{eq:fit-inf})
(see text).}
\end{figure}

\begin{widetext}  

\noindent 
\begin{table}
\caption{Estimates of the ground state energy per site $e_{\infty}^{N}$ for
the $N$-leg spin-$S$ Heisenberg ladders. The results in parentheses
are some of the best estimates known from the literature. Estimates
of the ground state energy per site of the infinite two-dimensional
system is also acquired by an extrapolation (see text). }

\centering{}%
\begin{tabular}{lllllc}
\hline 
$N$$\qquad$ & $S=\frac{1}{2}$ & $S=1$ & $S=\frac{3}{2}$ & $S=2$ & $S=\frac{5}{2}$\tabularnewline
\hline 
\hline 
1 & -0.4431471 & -1.40148403897  & -2.828337  & -4.761248 & -7.1924\tabularnewline
 & ( -0.4431471...) {[}Ref. \onlinecite{troyer-nleg}{]}$\quad$ & (
 -1.4014840389) {[}Ref. \onlinecite{whitespin-1}{]}$\quad$ & ( -2.82833)
 {[}Ref. \onlinecite{hallbergetal}{]}$\quad$ & ( -4.7612481)
 {[}Ref. \onlinecite{Schollwockspin-2}{]} & ( -7.19223) {[}Ref. \onlinecite{ng2}{]} \tabularnewline
2 & -0.578043140180  & -1.878372746  & -3.930067  & -6.73256  & -10.2852\textcolor{red}{{} }\tabularnewline
 & ( -0.57802) {[}Ref. \onlinecite{troyer-nleg}{]}$\quad$ &  &  &  & \tabularnewline
3 & -0.600537 & -2.0204 & -4.2718 & -7.3565 & -11.274\tabularnewline
 & ( -0.60063) {[}Ref. \onlinecite{troyer-nleg}{]}$\quad$ &  &  &  & \tabularnewline
4 & -0.618566 & -2.0957 & -4.446 & -7.669 & -11.76\tabularnewline
 & (-0.61873) {[}Ref. \onlinecite{troyer-nleg}{]}$\quad$ &  &  &  & \tabularnewline
5 & -0.62776  & -2.141 & -4.553 & -7.865 & -12.08\tabularnewline
 & (-0.62784) {[}Ref. \onlinecite{troyer-nleg}{]}$\quad$ &  &  &  & \tabularnewline
6 & -0.6346 & -2.169 & -4.60 & -7.94 & -12.1\tabularnewline
 & ( -0.6351) {[}Ref. \onlinecite{troyer-nleg}{]}$\quad$ &  &  &  & \tabularnewline
$\infty$ & -0.6768 & -2.327 & -4.97 & -8.62 & -13.2\tabularnewline
 & ( -0.66931) {[}Ref. \onlinecite{Wiesesspin1-2-2d} {]}$\quad$ &  &  &  & \tabularnewline
\hline 
\end{tabular}
\end{table}

\end{widetext}

\subsection*{The spin gap: \textmd{$\Delta_{s}$}}

%Now let us investigate the spin gap excitation $\Delta=E_{0}(n)-E_{0}(0)$
%from the singlet ground state to the lowest excited state with $S_{total}^{z}=n$,
%where $n=S+1$ if $N$ is odd and $S$ is integer, and $n=1$ otherwise.
Let $E_{0}(S_{tot}^z)$ be lowest energy in the sector $S_{tot}^{z}=\sum_{i,j}<S_{i,j}^{z}>$.
The spin gap is given by $\Delta=E_{0}(n)-E_{0}(0)$, where $n=S+1$
{[}$S+1/2]$ if $N$ is odd and $S$ is integer {[}semi-integer{]},
and $n=1$ otherwise.
By using this definition we obtain the correct spin excitation associated
with the spin gap, as we explain in the following. 

It is well known
that the ground state of Heisenberg chains ($N=1)$ with \emph{integer
spins} can be understood by the valence bond solid (VBS) picture.\citep{AKLT}
Open spin-$S$ chains that are described by VBS states have effective
$S_{end}=S/2$ spins at each edge. Due to this fact the ground state,
in the thermodynamic limit, is $(S+1)^{2}$-fold degenerate. Indeed,
this degeneracy has been observed in open spin-$S$ Heisenberg chains
(see for example Ref. \onlinecite{kenspin1}). Due to this fact, the
mass gap for the \emph{integer} spin-$S$ chains, under open boundary
condition, must be calculated from the singlet ground state to the
lowest excited state in the sector $S_{tot}^{z}=S+1$ {[}or equivalently,
$E_{0}(S+1)-E_{0}(S)]$. This degeneracy can be understood by the
edge spins that in the thermodynamic limit do not interact with each
others.\citep{kenspin1} 

By analyzing the topological term of the non-linear sigma model Ng
in Ref. \onlinecite{ng} proposed that spin-$S$ chains
with \emph{semi-integer spins }also present edge spins of magnitude
$S_{end}=(S-1/2)/2$. Theses edge spins indeed have been observed.\citep{qin-edge,ng2,eggert}
Due to this fact, the spin gap excitation of chains with semi-integer
spin that must be investigated is $\Delta=E_{0}(S+1/2)-E_{0}(0).$

\begin{figure}
\begin{centering}
\includegraphics[width=7cm]{fig4a.eps}
\par\end{centering}

\begin{centering}
\includegraphics[width=7cm]{fig4b.eps}
\par\end{centering}

\begin{centering}
\includegraphics[width=7cm]{fig4c.eps}
\par\end{centering}

\begin{centering}
\includegraphics[width=7cm]{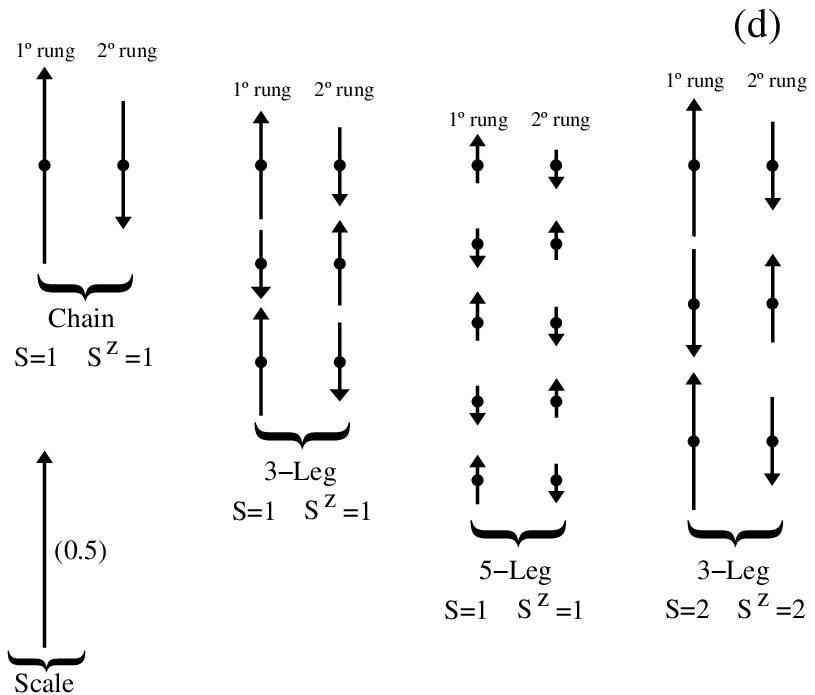}
\par\end{centering}

\caption{(color online) (a) A log-linear plot of $2\mid<S^{z}_{1,j}>\mid/S$
for chains with $L=400$ and $S=1$ and 2. Inset: a linear-linear
plot for three values of $S$. (b) $2N\mid<S^{z}_{1,j}>\mid/S$ vs $j$
for some ladders of size $L=60$ with spins 1 and 2 (see legend).
(c) $2N\mid<S^{z}_{1,j}>\mid/(S-1/2)$ vs $j$ for ladders with spins 3/2 and 5/2.
Only few sites are presented. (d) Values of $<S^{z}_{i,j}>$ measured in the
two first rungs of ladders with integer spins.  The size of the arrows indicates
the magnitude of $<S^{z}_{i,j}>$. The scale used is also presented. }
\end{figure}

The most beautiful signature of the edge states appears in the measured value
of $<S^{z}_{1,j}>$ in the sector $S_{tot}^{z}=S$ $[S-1/2]$ for integer
[semi-integer] spins. 
Let us focus, first, in  the results of integer spins.
%For the sake of clarity,
We show in Figs. 4(a) the local values of $<S^{z}_{1,j}>$ for \emph{chains}
%of size $L=400$ and
 with $S=1$,2 and $S=3$ (similar results can be found
in Refs. \onlinecite{white2} and \onlinecite{qin-edge} for $S=1$
and 2). It is clear from this figure, that a $S_{end}\cong S/2$ spin
appears at the end of the spin-$S$ chains. We found $S_{end}=0.53$, 
1.12, and 1.64 for the chains with $S=1$, 2 and 3, respectively.
As we can see from this figure, the log-linear plot shows an exponential
decay. The decay lengths $\xi^{S}$ obtained by the slope of straight
lines are: $\xi^{1}=5.94$ and $\xi^{2}=47.5$. The decay length $\xi^{1}$
is very close to the one found by White in Ref. \onlinecite{white}
($\xi^{1}=6.03$). On the other hand, our result for $\xi^{2}$ differ
from two previous estimates ($\xi^{2}=49.1$ and $\xi^{2}=54.3$,
see Refs. \onlinecite{Schollwockspin-2} and \onlinecite{qin-spin2b}).
It is also very interesting  to notice that in order to see the exponential
decay of $<S^{z}_{1,j}>$ from the edge, the size of the systems can
not be smaller than the decay length $\xi^{S}$ {[}see inset of Fig.
4(a){]}. Our result for the case $S=3$ {[}inset of Fig. 4(a){]} shows
that even considering a chain of size $L=800$ we were not able to
see yet an exponential decay. For chains with semi-integer spins
our results  support that  $S_{end}=(S-1/2)/2$, as expected.\citep{qin-edge,ng2,eggert} 
We found that $S_{end}=0.57$ and 0.94 for chains 
with spin S=3/2 and S=5/2, respectively. In those latter cases, we found
that $<S^{z}_{1,j}>$ exhibits a power-law decay [see Fig. 4(c)], 
as expected for open critical systems.

A similar effect that resembles what happens in 
chains also appears for ladders  
%with integer spins 
\emph{only} if $N$ is odd, as depicted
in Figs. 4(b) and (c) for integer and semi-integer spins, respectively. 
However, in this latter case the end spin
$S_{end}=\sum_{j=1}^{[\frac{N}{2}]}<S^{z}_{1,2j-1}>$ (where 
$\left\lfloor  \frac{a}{k} \right\rfloor$) 
is the largest integer less than or equal to $\frac{a}{k}$):
(i) is spread along the end rung {[}see Fig. 4(d){]} and (ii) the value of $S_{end}$
decreases with  the increasing of $N$.
%compared with the one of chains decrease with with increasing with the increasing number of  $N$.
%, i. e. $S/2$, decrease with the value
%of $N$. 
For example for $S=1$, we found that $S_{end}=0.45$, 0.33,
0.22 and 0.19 for $N=3$, 5, 7 and 9, respectively. For $N$ even
our results show the absence of this edge effect, as shown in Figs.
4(b) and 4(c). 
%for two representative cases. 

We can use a heuristic argument
to understand the fact that the edge states of ladders 
%with integer spins 
appear only for $N$ odd. Consider the strong coupling limit,
i. e., the coupling along the rung is much larger than the coupling
along the chains. If the number of legs is even (odd) the ground state
of the rung is a singlet ($S$-plet). So, ``rung sites'' behave
as an effective spin-$S$ if $N$ is odd. For this reason, we may expect
that $S_{end}=0$ for  $N$ even and $S_{end}=S/2 $ [$S_{end}=(S-1/2)/2$]  for
integer [semi-integer] spins and  $N$ odd. Indeed, our
results presented in Figs. 4(a)-(c) support this picture {[}see also
Figs. 6(a)-(c){]}.

%\begin{widetext}  
\begin{figure}[t]
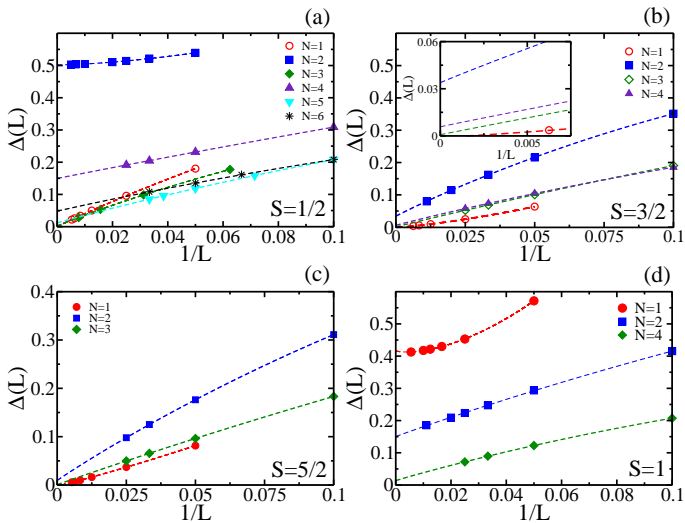

\begin{centering}
\includegraphics[width=4.5cm]{fig5a.eps}\includegraphics[width=4.5cm]{fig5b.eps}
\par\end{centering}
\begin{centering}
\includegraphics[width=4.5cm]{fig5c.eps}\includegraphics[width=4.5cm]{fig5d.eps}
\par\end{centering}

%\begin{centering}
%\includegraphics[width=5.5cm]{fig5d}\includegraphics[width=5.5cm]{fig5e}\includegraphics[width=5.5cm]{fig5f}
%\par\end{centering} 
\caption{(color online) (a), (b), (c) and (d) display the finite-size gap $\Delta(L)$
as a function of $1/L$ for the $N$-leg spin-S Heisenberg ladders
with spins $S=1/2$, 3/2, 5/2 and $S=1,$ respectively.  The symbols are the
numerical data and the lines in 
 these figures connect the fitted points (see text). Insets: zoom of the
region close to zero.}
 
\end{figure}
%\end{widetext}  

Finally, let us  show our results for the spin gap. In Figs. 5(a)-(d),
we present the finite-size spin gap $\Delta(L)$ as a function of $1/L$
for ladders with $S=1/2$, 3/2, 5/2 and 1 and some values of $N$. 
In order to estimate the spin gap in the thermodynamic limit $(\Delta_{s})$
we assume that $\Delta(L)$ behaves as

\begin{equation}
\Delta(L)=\Delta_{s}+\frac{A}{L}+\frac{B}{L^2}.\label{eq:fit-gap}
\end{equation}

It is expected that the spin gap of open chains with integer
spins behaves as\citep{afflecks1bosecond} $\Delta(L)=\Delta_{s}+\frac{B}{L^2}$.
Indeed, previous studies of the spin-1 chain\citep{white,afflecks1bosecond,qinspin-1-2}
found that $\Delta(L)$ scale with $1/L^{2}$ for large values of $L$.
We also have observed this behavior for the spin-1 chain {[}see Fig.
5(d){]}). We have added the $1/L$ term because our results show that the
leading finite-size correction of $\Delta(L)$ is $1/L$  for small system sizes.
Note that if we consider the energy dispersion 
$\Delta(k)=\sqrt{\Delta_{m}+(vk)^2}$ for a magnon with wave vector $k$, 
as point out by S. Qin $et$ $al.$ in Ref. \onlinecite{qinspin-1-2}, 
we see that $\Delta\sim\Delta_m+\frac{2\pi^2v^2}{L^2\Delta_m}$ only if
$L\gg\Delta_m$. 
This suggests that 
leading finite-size correction of $\Delta(L)$ is $1/L$ when the 
lattice sizes are smaller than the correlation length.
Even though the asymptotic scaling form is not reached, 
it is possible to obtain reasonable estimates of $\Delta_{s}$
by using Eq. (\ref{eq:fit-gap}), as we explain in the following.
It is possible to fit our data by using only the $1/L$ ($1/L^{2}$) term in
order to obtain a lower (upper) bound and then quote an estimative 
of the spin gap by an average of these two values, 
as done by Schollw\"ock and Jolicoeur in Ref. \onlinecite{Schollwockspin-2}. 
We notice that if we fit our data with the two terms, simultaneously, 
we get estimates very close to the procedure used by Schollw\"ock and
Jolicoeur, i. e., an estimate between the upper and lower bound, 
For example, for the spin-2 chain we got $\Delta_{s}=0.08$4 by considering $200<L\le400$
(it is interesting to mention that it was found $\Delta_{s}=0.07\pm02$
in experimental realization of the spin-2 chain\citep{exp-s2}). 
If we had considered systems with sizes $20<L<120$
to extract the spin gap, we would find $\Delta_{s}=0.04$.
This shows that if we consider system sizes smaller than the correlation length to the determine $\Delta_{s}$.
The estimate obtained should be interpreted as a lower bound estimate. 
Finally, we should mention that in principle 
logarithmic corrections like $1/L\ln L$
are also expected due to the marginally irrelevant operators.\citep{spinsaffleck}
However, we notice that if we replace the last term of Eq. (\ref{eq:fit-gap}) 
by the logarithmic correction term $1/L\ln L$,
the fits obtained are slightly worse than the ones found using the
Eq. (\ref{eq:fit-gap}).

%\begin{widetext}  
\begin{figure}
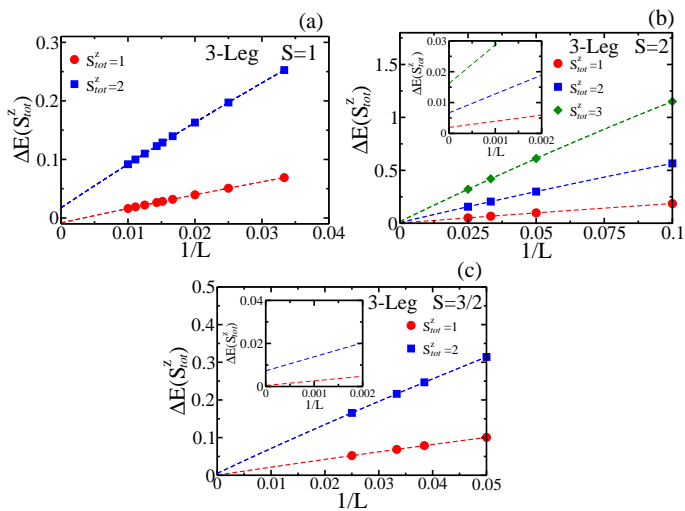

%\begin{centering}
%\includegraphics[width=5.5cm]{fig5a}\includegraphics[width=5.5cm]{fig5b}\includegraphics[width=5.5cm]{fig5c}
%\par\end{centering}

\begin{centering}
\includegraphics[width=4.5cm]{fig6a.eps}\includegraphics[width=4.5cm]{fig6b.eps}
\par\end{centering}  
\begin{centering}
\includegraphics[width=4.5cm]{fig6c.eps}
\par\end{centering} 
\caption{(color online)  (a), (b) and (c):
the spin excitation $\Delta E(S_{tot}^z)$ as function of $1/L$ for the 3-leg
Heisenberg ladders for $S=1$, $S=2$, and $S=3/2$ and some values of $S_{tot}^z$ (see
legend). The symbols are the numerical data and the lines in  these
figures connect the fitted points (see text). Insets: zoom of the
region close to zero.}
 
\end{figure}
%\end{widetext}  

As we already mentioned, it is expected that the $N$-leg spin-$S$
Heisenberg ladders is gapless (gapped) if $SN$ is semi-integer (integer).
Indeed, as we can observe in Figs. 5(a)-(d) our results, overall,
are consistent with the Haldane-Sénéchal-Sierra conjecture. We found
that for ladders with $SN$ semi-integer that the extrapolated values
of $\Delta_{s}$ are $\lesssim10^{-3}$. These latter results strongly
indicate that the spin gap is zero for ladders with $SN$ semi-integer.
In order to better visualize the results of $\Delta_{s}$, we report
in Table III the extrapolated values we got from the fit procedure.
In this table, we also present some  estimates of $\Delta_{s}$
found in the literature. As we can see, our results are similar to
those found in the literature and, within our precision, our results
agree perfectly with the Haldane-Sénéchal-Sierra conjecture. Note
that the spin gap decreases with the number of legs and the values of
the spins. Indeed, this bahavior is expected: by using the non-linear
sigma approach Sierra in Ref. \onlinecite{Sierralegs} (see also
Ref. \onlinecite{Chakravarty}) showed that the spin gap behaves as 
$\Delta_s\sim N S^2\exp (-SN a)$, where $a$ is constant.

%also that the spin gap decreases with the number of legs, as expected
%from the semi-classical approach.\citealp{Sierralegs} (see also Ref. )

As we discussed earlier, due to the edge states we expect that the
ground state is $(S+1)^{2}$-fold degenerate [$(S-1/2)^{2}$-fold degenerate], in the thermodynamic
limit, for ladders with integer [semi-integer] spins if $N$ is odd. In order to
verify this claim, we also calculate the spin excitations 
$\Delta E(S_{tot}^z)=E_{0}(S_{tot}^z)-E_{0}(0)$, for few values of $S_{tot}^z$.
%where $E_{0}(k)$ is lowest energy in the sector $S_{total}^{z}=k$.
In Figs. 6(a)-(c), we show the spin excitation $\Delta E(S_{tot}^z)$ for
3-leg ladders with spins $S=1$, 2 and $3/2$. The results for the 3-leg spin-1
Heisenberg ladder, presented in Fig. 6(a), in fact indicate that $E_{0}(1)=E_{0}(0)$
in the thermodynamic limit, as we expect. For larger values of $S$,
as we illustrate in Fig. 6(b)-6(c) for the cases $S=2$ and $S=3/2$, 
it is very difficult to see accurately if the ground state is degenerate (due mainly to
the system sizes we consider). Nevertheless, our results within of
the accuracy of the extrapolations are consistent with the fact that the ground state
is $(S+1)^{2}$-fold degenerate [$(S-1/2)^{2}$-fold degenerate] for ladders
with integer [semi-integer]    spins and
$N$ odd. Besides that, if we had estimated the spin gap for ladders
with integer spins considering only $\Delta E(1)$, we would find
that the spin gap for $N=4$ would be larger than $N=3$, which is
not expected. For these reasons we do believe that the ground state
is degenerate,  in the thermodynamic limit, for 
%integer spin
ladders with $N$ odd. 

\begin{widetext}  

\begin{table}[t]
\caption{Estimates of the spin gap $\Delta_{s}$ for the Heisenberg ladders
with up to six legs and $S\leq\frac{5}{2}$. These values were extracted
from the fit of our data using the equation (\ref{eq:fit-gap}). The
results in parentheses are some of the best estimates known from the
literature. }
%%%%
%%%%%
\centering{}% 
\begin{tabular}{lllllc} \hline $N$$\qquad$ & $S=\frac{1}{2}$ &
             $S=1$ & $S=\frac{3}{2}$ & $S=2$ & $S=\frac{5}{2}$\tabularnewline \hline \hline 
1 & --- & 0.41025 & --- & 0.084 & ---\tabularnewline   &  \hspace*{2.3cm} &
(0.41050) {[}Ref.  \onlinecite{whitespin-1}{]}$\quad$ & \hspace*{2cm} &
(0.085) {[}Ref.  \onlinecite{Schollwockspin-2}{]} & \tabularnewline 
2 & 0.5011 & 0.151 & 0.036 & 0.013 & 0.01\tabularnewline   &  ($0.504)$
{[}Ref. \onlinecite{whitenleg}{]} & \hspace*{2.3cm} &  &   & \tabularnewline 
3 & --- & 0.017 & --- & 0.01 & ---\tabularnewline   &  &  &  &  & \tabularnewline 
4 & 0.15 & 0.015 & 0.007 &  & \tabularnewline   & ($0.17$)
{[}Ref. \onlinecite{troyer-nleg}{]} &  & &  & \tabularnewline 
5 &  --- &  & --- &  & ---\tabularnewline   &  &  &  &  & \tabularnewline 
6 & 0.05 &  &  & & \tabularnewline   & (0.05) {[}Ref. \onlinecite{troyer-nleg}{]} &  & &  &
\tabularnewline \hline 
\end{tabular} 
\end{table}

\end{widetext}

\section{Conclusions}

In this work, we use the unbiased DMRG method to investigate the $N$-leg
spin-$S$ Heisenberg ladders. While the low energy physics of the
Heisenberg chains and the spin-1/2 Heisenberg ladders were studied
by several works, few is known about the Heisenberg ladders with spin
$S\ge1$. We made a great numerical effort to provide some precise
estimates of the ground state energy per site
$e_{\infty}^{N}$ in the thermodynamic limit for the Heisenberg ladders.
We also present several new estimates of the spin gap $\Delta_{s}$,
which were unknown, mainly for $N>1$ and $S\geq1$.
Our estimates for spin-$S$ chains and for spin-1/2 ladders are similar
to those known in the literature. 
%We also report several new estimates
%which were unknown, mainly for $N>1$ and $S\geq1$. 
Our results
corroborate with the Haldane and Sénéchal-Sierra conjectures for chains
and ladders, which establish that the $N$-leg spin-$S$ Heisenberg
ladders is gapless (gapped) if $SN$ is semi-integer (integer). We
also observe edge states for ladders if $N$ is odd that
resemble the edge states found in chains.
% with integer spin. 
We believe that this latter result will help understanding
 more deeply the nature of the ground state of the Heisenberg ladders. 
\begin{acknowledgments}
The authors thank G. Sierra and J. A. Hoyos for useful discussions.
This research was supported by the Brazilian agencies FAPEMIG and
CNPq. 
\end{acknowledgments}
\bibliographystyle{apsrev4-1}
%\addcontentsline{toc}{section}{\refname}\bibliography{/home/jcandido/FILES/textos/refs_rev4}
%

\end{document}